# SysMART Outdoor Services: A System of Connected and Smart Supermarkets

Yazan M. Mohamad, Majd G. Makdessi, Omar A. Raad, and Issam W. Damaj
Electrical and Computer Engineering Department
American University of Kuwait
Salmiya, Kuwait
{S00026193, S00027923, S00026922, idamaj}@auk.edu.kw

*Abstract*—Smart cities are today's modern trend. Many high-tech industrial firms are exploring different approaches to implement smart cities. Various projects aim at internet-of-things and smart solutions. Current implementations are mostly localized to a specific building or area; however, the growth is crossing space and geographic location limits. Shopping is a central activity that is frequent and typically a time-consuming task. SysMART is system of connected and smart supermarkets. SysMART enables a plausible shopping experience for customers. The aim of SysMART is to provide an advanced lifestyle with its ease of use functionality. SysMART outdoor services support distant parking availability, traffic status, and remote inventory checks for supermarkets in a chain. SysMART implementation relies on cutting edge technologies that support rapid prototyping and precision data acquisition, such as, National Instrument devices. The selected development environment is LabView with its world-class interfacing libraries. The paper comprises a detailed system description, development strategy, interface design, software engineering, and a thorough analysis and evaluation.

## I. Introduction

Smart cities are an important concept nowadays with growing potentials and need. Many projects aim at internet-of-things (IoT) and smart solutions but within single buildings or limited in geographical areas, such as, smart homes. Indeed, smart cities target larger populations and incorporate arrays of smart individual building management systems. The main purpose of smart cities is to increase quality, performance and interactivity of urban services, to reduce costs and resource consumption and to improve contact between citizens and government [1]. Many services and different aspects fall under the smart city concept, such as, government services, transportation and traffic management, energy, health care, and smart shopping [2]. Still, due to the wide aspect of smart city and the huge amount of projects that are implemented under this concept, it is hard to come up with a specific definition of a smart city.

The rise of new Internet technologies such as cloud-based services and use of smart phones, helped to solve problems in different and more accurate ways [3]. IoT is the network of physical objects embedded with electronics, software, sensors, and network connectivity, which enables these objects to collect and exchange data. The application of IoT in supermarkets and groceries are various and comprises indoor and outdoor services.

Supermarket outdoor services is usually focused on online shopping experiences, delivery, smart parking management, etc.

The benefits of different smart parking systems can be implemented for customer to save time looking for a parking. In [4], the author discusses how cities use regulations to minimize parking inside the city by issuing tickets and limiting the locations. In addition, [4] presents a smart parking implemented in Baltimore/Washington International airport in the United States (US) where each parking space is fitted with sensors to detect if it is occupied or not and thus informing the drivers of the available spaces. The system costs $450 per space; the cost is justified with a proof of efficiency that the system provides and the improved experience quality.

Zhou et al. [5] present an intelligent parking guidance system based on ZigBee network and geomagnetic sensors. The author tested in real time where the vehicles are positioned and information about the traffic was collected by geomagnetic sensors around parking lots and updated via ZigBee network. Furthermore, a Liquid Crystal Display (LCD) screen was used to show the number of available parking places. The experimental result proves that the distance detection accuracy of geomagnetic sensors was within 0.4 m, and data loss of wireless network in the range of 150 m is 0%. The system implemented in [5] is shown in Figure 1. In [6], an intelligent parking guidance system based on ZigBee network and geomagnetic sensors is proposed.

SysMART outdoor services provide advanced lifestyle in a smart city of connected supermarkets. The proposed system improves and modernizes the customer shopping experience. SysMART features unique off-supermarket options, such as, parking availability, roads to supermarkets traffic status, products availability, and provides an effective and user-friendly Internet interface. The customer uses SysMART Android app which access the database to check for the most convenient supermarket to visit based on traffic, parking spaces and product availability therefore enhancing the shopping experience.

This paper is organized so that Section II presents the system design, organization and architecture. Section III presents the system implementation. A thorough analysis and evaluation with a deployment example are presented in Section IV. Section V concludes the paper the sets the ground for future work.

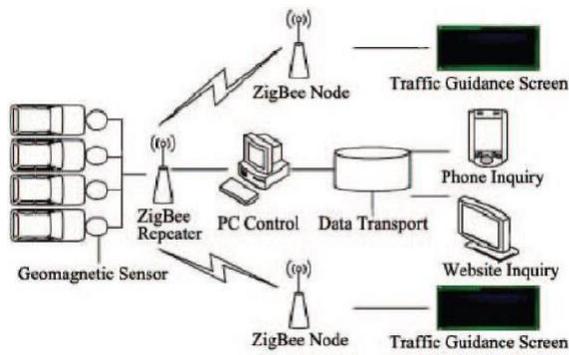

Fig 1. Parking guidance system

## II. SYSTEM ORGANIZATION AND ARCHITECTURE

SysMART outdoor services are ubiquitous and pervasive. The parking space and traffic status checkers and the item availability options are fully accessible online and connected to the Internet through an Access Point (AP) and a central server that hosts the services. The customer uses SysMART Android app to check the traffic status, parking spaces and product availability (See Figure 2). The digital hardware subsystem that implements the traffic status and parking space uses ultrasonic sensors connected to an Arduino Fio and an Xbee Series 2 for each node to communicate with a coordinator consisting of an Xbee with an Arduino processor. The Arduino processor updates the status regularly on the cloud-based database.

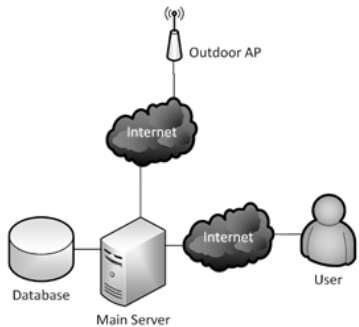

Fig 2. Outdoor System Architecture showing the outdoor access point (AP), the user, and the main server with its database.

SysMART outdoor services comprises three subsystems. In the parking and traffic subsystems, SysMART uses ZigBees with geomagnetic sensors to know whether a parking spot is available or not with the support of ultrasound sensors. The sensing information is transmitted to another ZigBee device that gathers that information to display to the customers how many spots are available (See Figure 3). SysMART services help in reducing congestion on roads, and parking spaces. For product availability checks, SysMART enables accesses to the supermarket chain databases for all branches.

## III. SYSTEM IMPLEMENTATION

The parking and traffic subsystems of SysMART are implemented using coordinators, nodes, the software interface, the database, and the mobile application. The system nodes consist of the following hardware:
- SparkFun Fio v3 Arduino-compatible interfacing board
- Ultrasonic sensor
- DC step-up shifter
- LEDs
- XBee Series 2
- Resistors
- Polymer Lithium Ion Battery

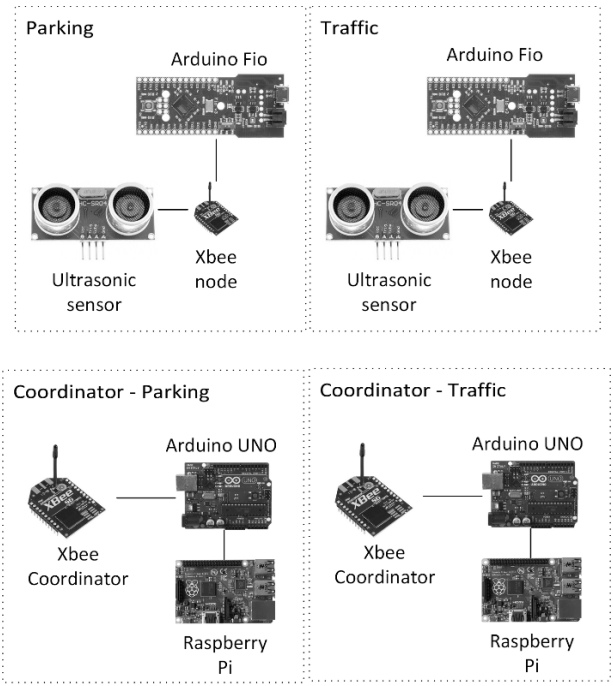

Fig 3. Traffic and Parking System Architecture Components

The ultrasonic sensor measures the distance from the nearest object in its range and sends that distance to the Sparkfun Fio. According to the measured distance, the XBees on the nodes send the status to the coordinator, so if there is a car in the parking spot, the node for this slot informs the coordinator and vice versa. Moreover, for the parking nodes there are two LEDs red and green that represent the status of the parking if the parking is busy the red led turns on otherwise the green led turns on. The system coordinators consist of Arduino UNO, XBee shield, XBee Series 2 module, and a Raspberry Pi processor. The coordinator communicates with all the nodes through the XBee network, the coordinator for the parking gets data from each parking slot to know their status, also, the coordinator for the traffic gets the data from several nodes placed near the road. Each coordinator has its Arduino UNO that collects the data coming from the nodes through the XBee. Finally, the Raspberry Pi processor receives the data from the Arduino UNO processor and updates it on the database through a Wi-Fi connection.

SysMART hardware subsystem is connected to the database which is designed for the connected smart supermarkets. The main entity in the database is the Store table which has the connected stores with their main information as shown in Table 1. The Store table is accessed by the mobile application to display for the stores, their parking slot status, and the traffic status around the stores. Another entity in the database is the inventory table which has the data for the connected stores representing the available products in each store and their quantity (See Table 1).

Table 1. Store Table of the database presented at the store

| Store _id | Store_ name | Store_l ong | Store_ lat | Store_parking _total | Store_parking_ available | Avg_tra ffic |
|---|---|---|---|---|---|---|
| 1 | Sultan _salmi yah | 29.342 31300 | 48.07 51530 0 | 3 | 2 | 2 |
| 2 | Sultan _shaab | 29.340 97700 | 48.04 42170 0 | 3 | 3 | 3 |
| NULL | NULL | NULL | NULL | NULL | NULL | NULL |

Table 2. Inventory table of the database presented at the store

| Store_id | Product_id | Product_location | Availability_in_store | Price |
|---|---|---|---|---|
| 1 | 1 | 1 | 5 | 2.000 |
| 1 | 2 | 2 | 4 | 0.100 |
| 1 | 3 | 3 | 1 | 1.000 |
| 1 | 4 | 4 | 5 | 5.000 |
| 1 | 5 | 4 | 1 | 1.000 |
| 1 | 6 | 6 | 6 | 6.000 |
| NULL | NULL | NULL | NULL | NULL |

Each supermarket has one coordinator that waits until it receives the data from all nodes or until a timeout occurs, accordingly the parking availability is updated.

Android studio is used to develop SysMART application. The application is used outdoor when the customer is away from the supermarket and wants to find information regarding the products offered by the supermarket. The android application is split into 2 subparts, which are choose a store, and search items in all stores as shown in Figure 4(a). The application is designed to be user-friendly. In the mobile application, choosing a store shows a list of all stores. The customer needs to pick a store.

Then the application shows the parking availability in the chosen store, and traffic status in the surrounding area of the store. In addition, search items in the store is another option that can be selected; it is used to check the availability of the products. First, the lists are supplied showing all the items that are available at the store. The second feature is search items in all stores, it shows the customers a list of categories to choose the items from. Then, a table displays the availability of the products selected by the customer and in all stores. The navigation trace of the application is shown in Figure 4.

## IV. EVUALATION AND ANALYSIS

SysMART outdoor hardware subsystem includes sensor nodes and coordinators with a variety of characteristics. Each sensor node has an ultrasonic sensor with power consumption of 75 mW, an Xbee series 2 with 132 mW, and Arduino Fio with 25 mW, and a DC-DC step up with 1 mA. The maximum power consumption for each node is in the order of 230 mW when active and 50 mW when idle or sleep. For the coordinator, it uses a Raspberry Pi 3 with 10000 mW and an Xbee Series 2 with 132 mW, making the total power consumption 10132 mW. As the coordinator must remain active at all times, for a stable communication, the power consumption cannot be reduced. Moreover, each sensor node has a range of 400 ft. The range is extended by using the routing feature of the Xbee to route the packets among nodes and the coordinator for a wider coverage. The Xbee communicates 21-byte packet at 250 Kbps in 672 microseconds and the node sends an update every 500 milliseconds. At the coordinator side, the Raspberry Pi 3 is connected to the supermarket's access point over WiFi using 802.11g which support 54 Mbps communication.

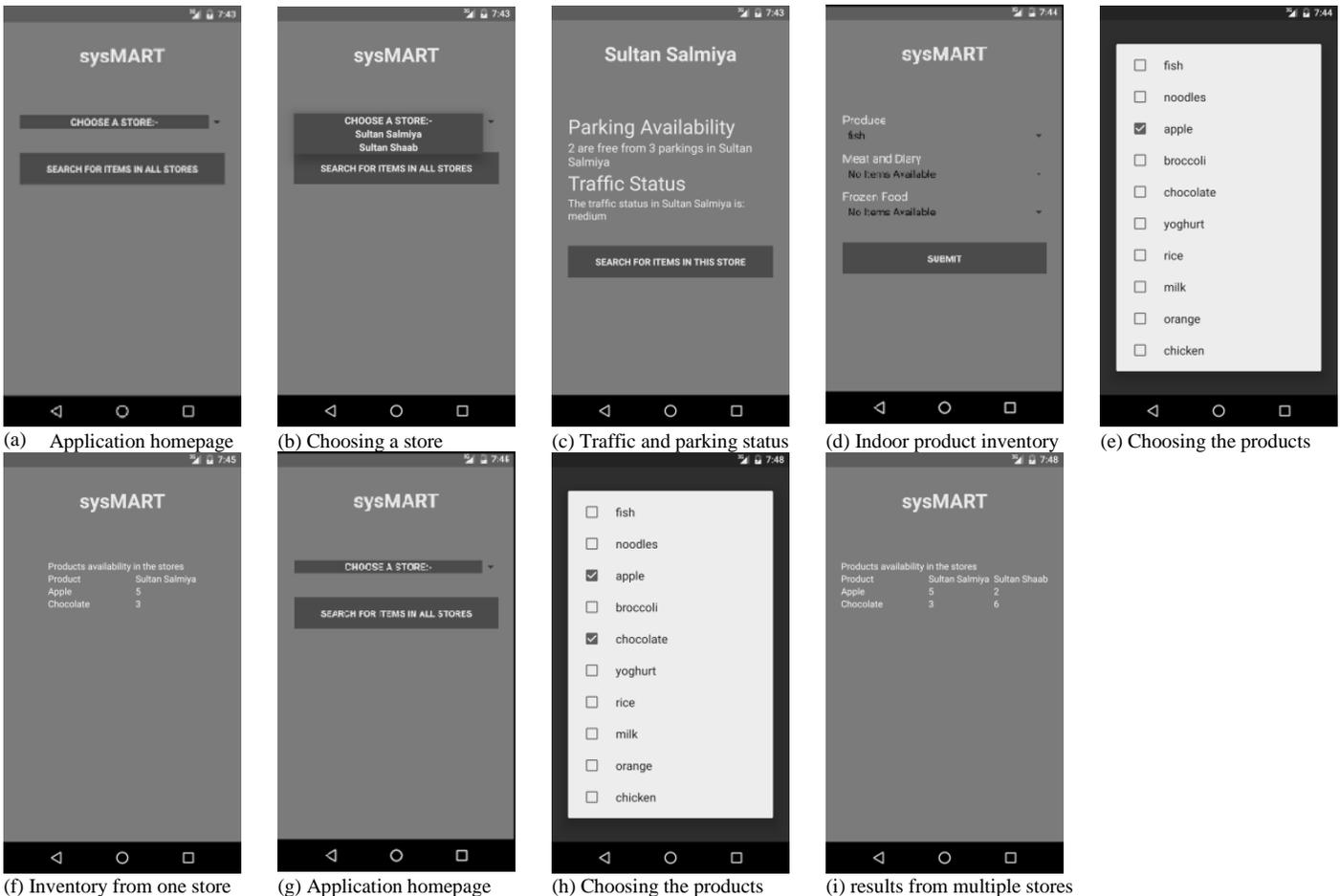

(a) Application homepage  (b) Choosing a store  (c) Traffic and parking status  (d) Indoor product inventory  (e) Choosing the products

(f) Inventory from one store  (g) Application homepage  (h) Choosing the products  (i) results from multiple stores

Fig 4. Application sample use

SysMART impacts the society in different aspects. A survey has been conducted to evaluate the society point of view towards the features offered in SysMART. The survey was given to 58 candidates of which 27 people were between the ages of 18 to 24, eight people were between the ages of 25 to 34, ten people were between the ages of 35 to 44, ten people were between the ages of 45 to 54, and three people were between the ages of 55 to 64.

The survey includes the question "*if a supermarket that uses smart technologies to ease the customer experience would be in demand*". 48 people answered yes, it will and they will visit that supermarket more often. The candidates were also asked "*if a mobile application that tells you the parking availability, traffic status near the grocery store, guides you to your product, tells you the fastest checkout lane, and tracks the temperature of the products at all times helps*". Almost 85% of the candidates answered yes. Furthermore, the candidates were asked to rate the features that SysMART presents. From the results, the SysMART features are accepted and given an average rating of around 4.5 out of 5 being the best (See Figure 5).

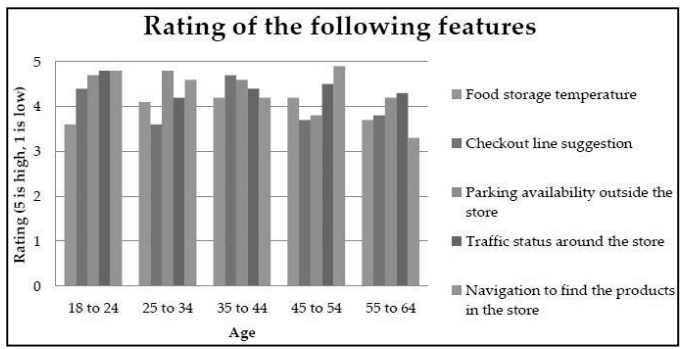

Fig 5. Questionnaire results

To evaluate SysMART, a deployment study is applied on The Sultan Center Chain for one branch. The Sultan Center is a chain of local supermarket stores available mainly in Kuwait. The Salmiya area branch is located between two main roads; it has two parking spaces facing each street. Figure 6 shows the supermarket satellite image and identified using a while rectangle. Deploying the outdoor subsystem, the supermarket needs 8 traffic sensor, 4 for each street, 2 for each direction located at opposite sidewalks. Moving to the parking spaces the first parking space can hold 70 cars and the second parking can hold about 20 cars so in total the supermarket is using 90 parking sensors. Kuwait is mostly sunny throughout the year with the parking spaces available outdoor so we can power the sensor s using solar power. Table 3 details the budget needed to implement SysMART's outdoor system for the selected Sultan Center branch. The estimated total is around 10,000 USD.

## V. CONCLUSIONS

SysMART aims to enhance the shopping experience and save the customers' time. The project implementation relies on cutting edge technologies that support rapid prototyping and precision data acquisition. SysMART supports the shopping starting from home till check out from the store by providing several options for the customer through a mobile application that connects supermarkets together. The user can check the availability of the items that one is willing to purchase in specific supermarket, or has the option to let SysMART suggest alternatives as per product availability, parking readiness, and traffic status. The parking spot availability and traffic status is updated according to specific hardware connected to the server. SysMART can be improved by supporting reserving the parking slot online, and add graphical maps to support the parking status checking. Moreover, the system can be developed by improving the hardware to decrease the cost and the power consumption.

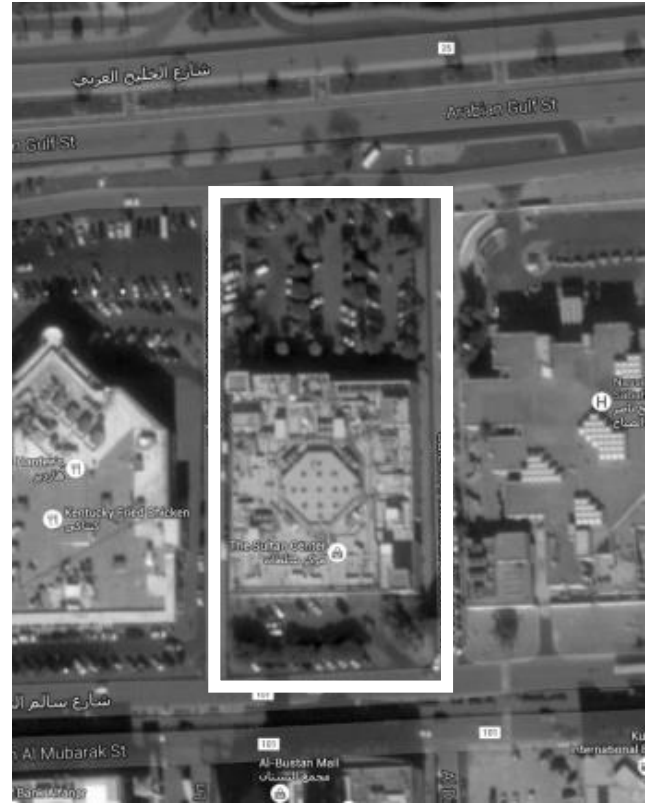

Fig 6. The Sultan Center Salmiya branch satellite image is provided by Google Maps

Table 3. Cost to apply SysMART for The Sultan Center branch

|  | Cost/Unit | Qty | Cost |
|---|---|---|---|
| Arduino Fio | $35 | 98 | $3,430 |
| Xbee Series 2 | $23 | 100 | $2,300 |
| Ultrasonic Sensor | $2 | 98 | $196 |
| Battery | $13 | 98 | $1,274 |
| Solar Panel | $6 | 98 | $588 |
| Raspberry Pi 3 | $35 | 2 | $70 |
| Software | $1,000 | 1 | $1,000 |
| Maintenance/year | $1,500 | 1 | $1,500 |
| Arduino UNO | $25 | 2 | $50 |
|  |  |  |  |
| Total |  |  | $10,408 |